\documentclass[10pt,letterpaper]{article}
\usepackage[top=0.85in,left=2.75in,footskip=0.75in,marginparwidth=2in]{geometry}

\usepackage[utf8]{inputenc}

\usepackage{cite}
\usepackage{amsmath}

\usepackage{nameref,hyperref}
\usepackage[right]{lineno}

\usepackage{microtype}
\DisableLigatures[f]{encoding = *, family = * }

\raggedright
\usepackage{changepage}
\usepackage[aboveskip=1pt,labelfont=bf,labelsep=period,singlelinecheck=off]{caption}

\makeatletter
\renewcommand{\@biblabel}[1]{\quad#1.}
\makeatother

\usepackage{soul}
\usepackage{graphicx}
\usepackage{hyperref}
\usepackage{placeins}
\usepackage{mathptmx}
\usepackage{anyfontsize}
\usepackage{t1enc}
\usepackage{lastpage,fancyhdr,graphicx}
\usepackage{epstopdf}
\usepackage{cite} 
\usepackage{multirow}
\usepackage{url}
\usepackage{tabularx}
\usepackage{placeins}
\usepackage{multicol,lipsum}
\usepackage{rotating}
\usepackage{natbib}
\fancyheadoffset[L]{2.25in}
\fancyfootoffset[L]{2.25in}

\usepackage{color}

\definecolor{Gray}{gray}{.25}

\usepackage{graphicx}

\usepackage{sidecap}

\usepackage{wrapfig}
\usepackage[pscoord]{eso-pic}
\usepackage[fulladjust]{marginnote}
\reversemarginpar

\begin{document}
\vspace*{0.35in}

\begin{flushleft}
{\Large
\textbf\newline{Resonant absorption as a damping mechanism for the transverse oscillations of the coronal loops observed by SDO/AIA}
}
\newline
\\
Javad Ganjali\textsuperscript{1},
Nastaran Farhang\textsuperscript{1},
Shahriar Esmaeili\textsuperscript{2},
Mohsen Javaherian\textsuperscript{3}, and
Hossein Safari\textsuperscript{1}
\\
\bigskip
\bf{1} Department of Physics, University of Zanjan, University Blvd., Zanjan, 45371-38791, Iran,\\
\bf{2} Department of Physics and Astronomy, Texas A\&M University, 4242 TAMU, University Dr, College Station, TX 77840, USA,\\
\bf{3} Research Institute for Astronomy and Astrophysics of Maragha (RIAAM), Maragha, 55134-441, Iran.
\bigskip


\end{flushleft}
\justify
\section*{Abstract}
Solar coronal loops represent the variety of fast, intermediate, and slow normal mode oscillations. In this study,  the transverse oscillations of the loops with a few-minutes period and also with damping caused by the resonant absorption were analyzed using extreme ultraviolet (EUV) images of the Sun. We employed the 171 $\AA$ data recorded by Solar Dynamic Observatory (SDO)/Atmospheric Imaging Assembly (AIA) to analyze the parameters of coronal loop oscillations such as period, damping time, loop length, and loop width. For the loop observed on 11 October 2013, the period and the damping of this loop are obtained to be 19 and 70 minutes, respectively. The damping quality, the ratio of the damping time to the period, is computed about 3.6. The period and damping time for the extracted loop recorded on 22 January 2013 are about 81 and 6.79 minutes, respectively. The damping quality is also computed as 12. It can be concluded that the damping of the transverse oscillations of the loops is in the strong damping regime, so resonant absorption would be the main reason for the damping.

\textbf{Key words:} Sun: corona – Sun: magnetic fields – Sun: oscillations.

\section{Introduction}\label{sec1}
The field of coronal seismology has been developing during last decades. It seems to be heading toward revolution in the physics of the Sun. It means a very efficient instrument is achieved to explore the basic intrinsic physical parameters of the solar corona including magnetic field, temperature, and density of plasma. The slow, intermediate, and fast oscillations of the solar coronal loops were detected by various types of space telescopes \cite{aschwanden_1999, Nakariakov_1999, Ofman_tong_2002, wang2003, Wang_2004, Verth_2007, De_Moortel_2007, Safari_2007, erdelyi2008, Dadashi_2009, Verwichte_2009, ashwanden2011, Taran_2014,Abedini_2016, Abedini_2018}.

From the theoretical point of view, the coronal seismology was proposed by \cite{uchida} and \cite{roberts_1984} for the flux tube standing waves. The coronal seismology is based on the dispersion relation defined for a plasma cylinder; a cylinder with non-uniform plasma structure formed by a magnetic field \cite{edwin_1983, Goossens_2002, Nakariakov_1999, Ofman_2002, Ruderman_2002, Karami_2002, van_Doorsselaere_2004_a, van_Doorsseleare_2004_b, Andries_2005, Erdelyi_2007, Safari_2006, Abedini_2012, Esmaeili_2015,Esmaeili_2016, Esmaeili_2017, Vasheghani_Farahani_2017}.

With the advancement of technology, due to high-resolution spatial imaging, the new coronal seismological field entered the golden age of explorations. The first evidence for fast MHD kink mode was achieved from TRACE observation, which was based on detecting the transverse loop movement oscillations with the theoretically expected periods of kink mode \cite{Schrijver2002}. Some kink modes seem to be majorly in the transverse direction \cite{aschwanden_1999}; while other modes clearly oscillate in perpendicular direction \cite{Wang_2004}. 

The second-order geometrical and physical effects of coronal oscillation have been theoretically studied; nevertheless, the effect of curvature of loops on the oscillation period \cite{van_Doorsseleare_2004_b}, the impact of the elliptic transverse cross sections on damping of oscillation \cite{ruderman2003}, and also the effect of density stratification on the loop oscillation have been carefully specified \cite{Mendoza_Briceno_2004, Andries_2005, Dymova_2005, Safari_2007, Verth_2007, Fathalian_2010, Verth_2010, Soler_2011, Farahani_2014, Grant_2015, Pascoe_2017, Shukhobodskiy_2018, Pascoe_2018}.

The observation of the first two harmonics of the horizontally polarized kink waves excited in the coronal loop system was reported by \cite{Guo_2015}. \cite{Zhang_2016} also investigated the evolution of two prominences $(P_{1},P_{2})$ and two bundles of coronal loops $(L_{1},L_{2})$. Another development concerning the nondamping oscillations at flaring loops was recently published by \cite{Li_2018}. As a seismological application, periods and damping rates of the fast sausage oscillations in multishelled coronal loops were investigated by  \cite{Chen_2015}. Also, \cite{Jin_2018} recently studied the damping of two-fluid MHD Waves in stratified solar atmosphere.

Analysis of the transverse oscillations of loops (kink mode) in an active flaring region indicates that the initiation of these oscillations is caused by a disturbance, movements from the center of a flare towards the outside at the velocity of 700 km/s, which can produce a shock wave \cite{Aschwanden_2006}. The understanding of coronal loop oscillations and the mechanism underlying their damping has been subjected to a vast studies. Various damping mechanisms for oscillations of the coronal loops have been discussed by \cite{Roberts_2000}.

\cite{Hollweg_1988} were the first ones who investigated and discussed the damping of kink oscillations caused by resonant absorption. A method to analyze the dissipative processes in the regimes with the vicinity of the singularity has been developed by \cite{Sakurai_1991}, \cite{Sakurai_1991_b}, \cite{Goossens_1992}, and \cite{Goossens_1995}. \cite{Ruderman_2002} rebuilt this idea. They considered this problem for a straight magnetic flux tube disturbed in cold plasma. The tube had a homogeneous core and a thin layer of thickness $l$ whose density uniformly reduced from center to the tube boundary. 

Resonant absorption occurs when the waves entered the flux tube from the footpoints area are frequently reflected and the kink oscillation frequency of the tube becomes equal to the local Alfv\'{e}n frequency in a place within the resonant layer of the tube. Thus, resonantwill occur for the standing waves, and the energy of these waves will be converted to the thermal energy of the environment through the ohmic resisitivity and viscous dissipation. 

Considering the fact that the damping time caused by resonant absorption is about the order of $(a/l)P$, where $l$, $a$, and $P$ are length, radius, and period of the loop respectively, \cite{Ruderman_2002} employed a new proposed mechanism for the data observed by \cite{Nakariakov_1999}, and concluded that $l/a=0.23$. \cite{Goossens_2006} used the mechanism proposed by \cite{Hollweg_1988}, and \cite{Ruderman_2002} in order to estimate the amount of $(l/a)$ for 11 damped loops. They obtained this value in the range $0.16$ to $0.491$. These answers were obtained by assuming $l \ll a$. These results were inspired by \cite{van_Doorsselaere_2004_a} for elimination of the limitation of $l \ll a$ as well as numerically solving the damping of the loop. They came to the conclusion that the difference between numerical and analytical values for $l/a \leq 1/3$ is very small. Even for $l \simeq a$, the difference was not more than $0.25$. Recently, \cite{Su_2018} investigated the strength of the magnetic field using the densities obtained by the differential emission measure (DEM) method and they concluded that the magnetic field decays during the oscillation.

\cite{Goossens_1992} studied the resonant absorption and obtained the ratio of the damping rate to the oscillation frequency for the long wavelengths in case the magnetic field is constant and parallel to the axis of the tube everywhere. The ratio of the damping time to the oscillation period is obtained 4.97, which has a value of about 3 to 5 based on observations. Eventually, we conclude that the resonant absorption is an acceptable mechanism for explaining the damping observed in the transverse kink oscillations of coronal loops.

This paper is organized as follows, the method applied for extracting the oscillation is introduced in Section \ref{sec2}. In Section \ref{sec3}, we present the extracted results for frequencies and damping times of the loops recorded on 22 January 2013 and 11 October 2013. The main conclusions is also presented in Section \ref{sec4}.

\section{Method}\label{sec2}
The coronal loops of the Sun are curved and bright structures. The hot plasma trapped around the magnetic field lines inside the loop leads to seem brighter than their surrounding environment. Due to factors such as fast oscillating waves caused by flares, coupling with oscillating modes enforced by pressure, and pulses of driven hot plasma, these magnetic loops represent the normal oscillation modes. The coronal loops have lengths from a few mega meters to several hundred mega meters. The limitation of the spatial resolution capability of solar observatories in the EUV and X-ray pass bands has made it almost impossible to observe and analyze the internal structure of the thin loops \cite{Esmaeili_2016, Esmaeili_2017}. Coronal seismology provides an alternative method for understanding the physical and geometrical structure of the loops.

Pursuing higher spatial resolution, the Solar Dynamic Observatory (SDO) spacecraft was launched in February 2010 to study the Sun interior, solar magnetic field, solar coronal hot plasma, and effects of photospheric phenomena on space weather. In this study, we used the solar data provided by the Atmospheric Imaging Assembly (AIA) instrument on-board the SDO which provides full-disk images of the Sun's atmosphere with a time cadence of 12 seconds in various EUV pass bands. Therefore, the damping of loops are we investigated using successive EUV images at 171 \AA.

\begin{figure*}[ht]
\centerline{\includegraphics[width=1.3\textwidth,clip=]{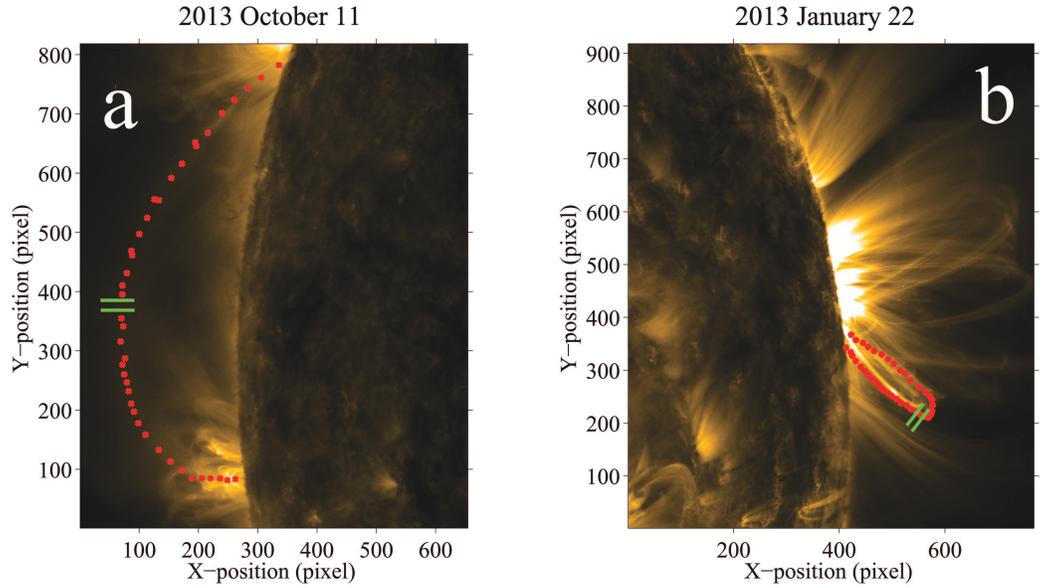}}
\caption{Partial image of the Sun at 171 {\AA} provided by SDO/AIA. Left panel: the observed loop on 11 October 2013. Right panel: the observed loop on 22 January 2013.}\label{fig1}
\end{figure*}

To this end, we will address the creation of space-time images from consecutive EUV images of the coronal loops at a specified time interval. By means of the Gaussian function fitting to each time element of these images, parameters such as the spatial oscillation amplitudes and the width of the loops are extracted. Then, through analyzing the spatial oscillation amplitudes of the loop, the periods and the damping time are obtained. In this study, consecutive images of loops on 11 October 2013 at 07:11:59 to 08:21:59 UT  (Figure \ref{fig1}, a) and 22 January 2013 at 02:20:00 to 03:41:00 UT (Figure \ref{fig1}, b) provided by http://jsoc.stanford.edu are investigated. To extract the oscillations for each data set, we applied the displacements correction due to the differential rotation of the Sun. To co-align the consecutive images to one reference, all data were derotated (see  \cite{Alipour_2015}).

In order to create the space time image, arbitrary number of points (based on length and width of the loop) are selected using the spline interpolation in two directions to form a rectangular region perpendicular to the loop axis (as shown with green lines in Fig. \ref{fig1}). Averaging over intensities of distinct pixels on different rows within the appointed box creates an element of the space time image for the desirable locations at a specific time. Performing the same procedure for each successive image results in the space time image (Fig. \ref{fig2}), in which the transverse oscillating mode is extracted by averaging over the intensities perpendicular to the loop axis.

\section{Results and Discussion}\label{sec3}
\begin{figure*}[ht]
\centerline{\includegraphics[width=1.3\textwidth,clip=]{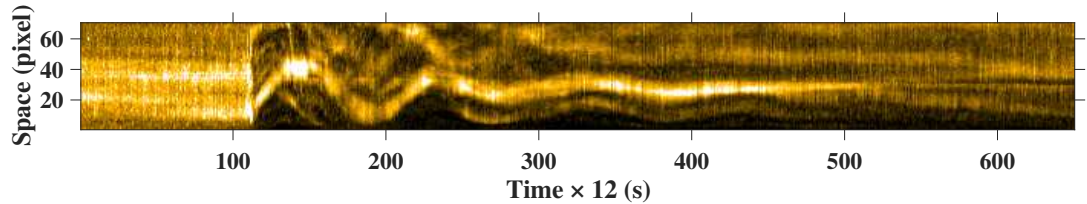}}
\caption{The space-time image of the loop observed on 11 October 2013, obtaining from 650 consecutive EUV images provided by SDO/AIA.}\label{fig2}
\end{figure*}

\begin{figure*}[ht]
\centerline{\includegraphics[width=1.3\textwidth,clip=]{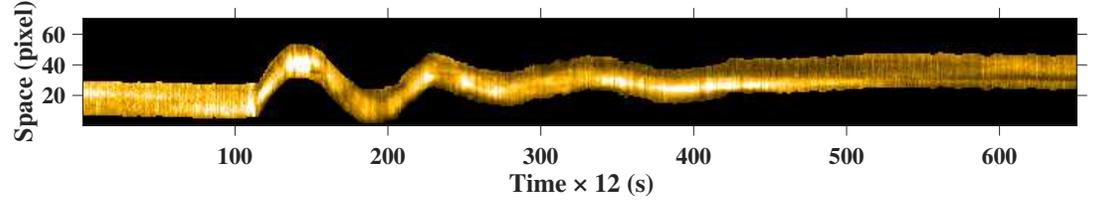}}
\caption{The extracted oscillating wave from the space-time image of Figure \ref{fig1}, corresponding to the loop observed on 11 October 2013.}\label{fig3}
\end{figure*}

In the space-time image of Fig. \ref{fig2}, the kink oscillation modes of several loops observed on 22 January 2013 are shown. Considering the complexity of the simultaneous analysis of all these oscillations, we just extract the most noticeable loop from this space-time image and address its parameters (Figure \ref{fig3}). This oscillations started at 07:11:59 UT and ended at 08:21:59 UT. A Gaussian function was employed to derive the oscillation amplitudes as follows:
\begin{equation}\label{eq1}
F(x, t) = f(t)\exp\left( -\left(\frac{x-a(t)}{\sqrt{2}\sigma(t)}\right)^2\right)+b(t).
\end{equation}
This fitting is performed for each column of the space-time image. In this regard, the parameter $f(t)$ is the index of intensity oscillation amplitudes, $a(t)$ is the index of spatial oscillation amplitudes, $x$ is the length of the space-time rectangle created in pixel unit, $\sigma(t)$ is related to the the width of Gaussian function, and $b$ represents the background intensity. The loop width is also obtained by $w=2\sigma \sqrt{2ln2}$. 

The lengths of the two studied loops are obtained about  $345 \pm 35$ Mm (Figure \ref{fig1}, a) and $195 \pm 20$ Mm (Figure \ref{fig1}, b), respectively. Therefore, using the values of $\sigma$, the width of loops are $6.85$ Mm and $3.7$ Mm, respectively.

\begin{figure*}[ht]
\centerline{\includegraphics[width=1.2\textwidth,clip=]{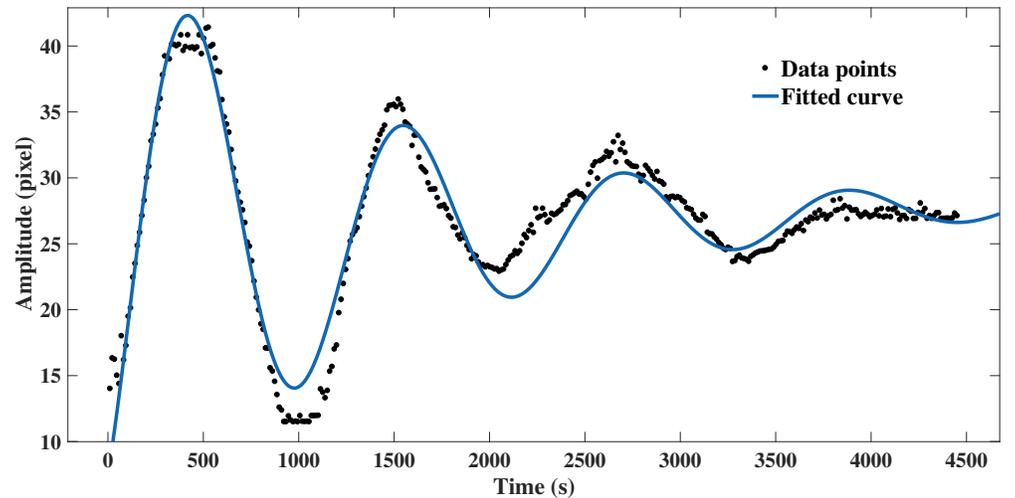}}
\caption{The spatial oscillation amplitudes of the loop recorded on 11 October 2013, and the result of fitting Eq. (\ref{eq2}) to the amplitudes.}\label{fig4}
\end{figure*}

At the next stage the period of studied loop is attainable by fitting the following function ($A(t)$) describing equilibrium position of the spatial oscillation amplitudes (e.g., \cite{Su_2018}):
\begin{equation}\label{eq2}
\begin{aligned}
   A(t) = a_0 + a_1 \exp \left( \frac{-(t-t_0)}{\tau} \right)\cos\left(\frac{2\pi (t-t_0)}{P_{0} + kt - \phi}\right) + a_2 \frac{t-t_0}{p_0},
\end{aligned}
\end{equation}
where $a_0$ is the amplitude, $t$ is time, and $\tau$ represents damping time. The parameters $P_{0}$, $k$, and $\phi$ are the period of oscillation, the evolution rate, and the oscillation phase, respectively. The spatial oscillation amplitudes of the loop recorded on 11 October 2013 and the corresponding fit are represented in Fig. (\ref{fig4}). 

According to the fit results, $\tau$ is obtained about 70 minutes and the oscillation period is obtained around 19 minutes. The damping quality, the ratio of the damping time to the period of oscillation {$(Q=\frac{\tau}{P})$} \cite{Safari_2006, Ruderman_2002}, is computed about 3.6. It seems that the main mechanism for the strong damping of the loop oscillation is the resonant absorption. 

\begin{figure*}[ht]
\centerline{\includegraphics[width=1.3\textwidth,clip=]{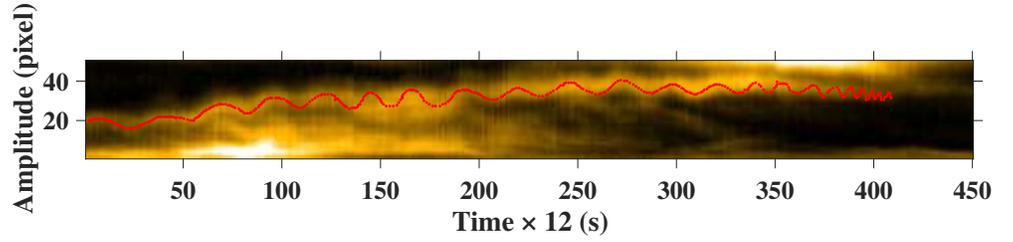}}
\caption{The space-time image belonging to the loop recorded on 22 January 2013 obtained from 450 consecutive EUV images provided by SDO/AIA.}\label{fig5}
\end{figure*}

\begin{figure*}[ht]
\centerline{\includegraphics[width=1.25\textwidth,clip=]{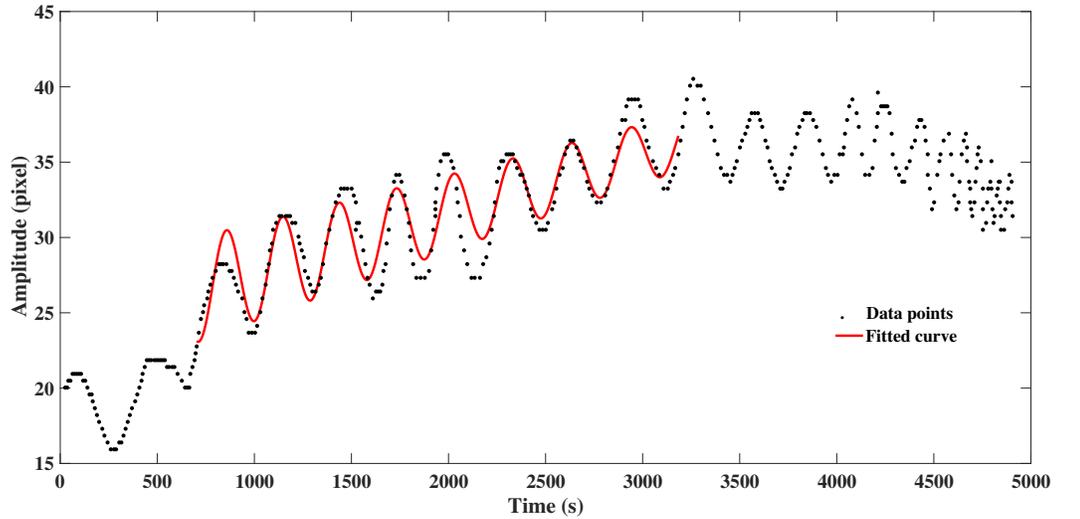}}
\caption{The spatial oscillation amplitudes of the loop recorded on 22 January 2013, and the result of fitting Eq. (\ref{eq2}) to the amplitudes over the interval 720 to 3182 seconds.}\label{fig6}
\end{figure*}

We have also studied the loop recorded on 22 January 2013 at 02:20:00 UT (Figure \ref{fig1}, b), following the same procedure described earlier. The oscillation of this loop started at 02:20:00 UT and ended at 03:41:00 UT. The the space-time image corresponding to this data set is shown in Figure \ref{fig5}. The period and damping time are obtained about 6.79 and 81 minutes, respectively. The damping quality is computed about 12. Figure (\ref{fig6}) represents the result of fitting Eq. (\ref{eq2}) to the spatial oscillation amplitudes of the loop over the interval 720 to 3182 seconds.

\section{Conclusions}\label{sec4}

In this paper, we studied the period and the damping time of transverse kink oscillations using successive EUV images for two loops recorded on  11 October 2013 (loop 1) and 22 January 2013 (loop 2) provided by SDO/AIA. The spatial oscillation amplitudes for each data set are derived using the appropriate Gaussian function  (Eq. \ref{eq1}). The lengths of the loop 1 and 2 are calculated $345 \pm 35$ Mm and $195 \pm 20$ Mm, respectively. We determined that the period and the damping time for the loop 1 are about 19 and 70 minute, respectively. For the loop 2, the period and damping time are also calculated about 81 and 6.79 minute, respectively.  Therefore, the damping quality for the loop 1 and 2 are 12 and 3.6, respectively. According to the obtained damping qualities for each studied case, which can be classified as the strong damping, we conclude that the resonant absorption may be the main mechanism for damping of these oscillations.

Several damping mechanisms (e.g., non-ideal MHD effects, lateral wave leakage, footpoints wave leakage, phase mixing and resonant absorption) were investigated for the damping of coronal loop oscillations (e. g. \cite{2005}) . The results of the present study and also many other previous research show that the kink mode oscillation of coronal loops are damped due to the resonant absorption in the strong damping regime. From the theoretical point of view (e. g., \cite{Ruderman_2002}, \cite{Safari_2007}), at a thin resonant layer (boundary layer) at the lateral boundary of the coronal loops with the inhomogeneous density along the cross section, the energy of the kink mode oscillations can be transferred to the localized Alfv\'{e}n waves. This energy may heat up the coronal loops to several million kelvins.

\FloatBarrier
\section*{Acknowledgements}
The authors thank NASA/SDO and the AIA science team for providing data publicly available.

\nolinenumbers

\bibliographystyle{rusnat}
\setcitestyle{authoryear}
\bibliography{library}

\begin{thebibliography}{60}
\providecommand{\natexlab}[1]{#1}
\providecommand{\EM}{\em}
\providecommand{\RNtxt}{\relax}
\RNtxt{}

\bibitem[Abedini(2016)A.~Abedini]{Abedini_2016}
{\EM Abedini A.}
\newblock Phase speed and frequency-dependent damping of longitudinal intensity
  oscillations in coronal loop structures observed with {AIA}/{SDO}
  \allowbreak\newblock// Astrophysics and Space Science. mar 2016. 361, 4.

\bibitem[Abedini(2018)A.~Abedini]{Abedini_2018}
{\EM Abedini A.}
\newblock Observations of Excitation and Damping of Transversal Oscillations in
  Coronal Loops by {AIA}/{SDO} \allowbreak\newblock// Solar Physics. jan 2018.
  293, 2.

\bibitem[Abedini et~al.(2012)A.~Abedini, H.~Safari, S.~Nasiri]{Abedini_2012}
{\EM Abedini A., Safari H., Nasiri S.}
\newblock Slow-Mode Oscillations and Damping of Hot Solar Coronal Loops
  \allowbreak\newblock// Solar Physics. jul 2012. 280, 1. 137--151.

\bibitem[Alipour, Safari(2015)N.~Alipour, H.~Safari]{Alipour_2015}
{\EM Alipour N., Safari H.}
\newblock {STATISTICAL} {PROPERTIES} {OF} {SOLAR} {CORONAL} {BRIGHT} {POINTS}
  \allowbreak\newblock// The Astrophysical Journal. jul 2015. 807, 2. 175.

\bibitem[Andries et~al.(2005)J.~Andries, M.~Goossens, J.~V. Hollweg,
  I.~Arregui, T.~van Doorsselaere]{Andries_2005}
{\EM Andries J., Goossens M., Hollweg J.~V., Arregui I., Doorsselaere T. van}.
\newblock Coronal loop oscillations \allowbreak\newblock// Astronomy {\&}
  Astrophysics. jan 2005. 430, 3. 1109--1118.

\bibitem[{Aschwanden} et~al.(1999)M.~J. {Aschwanden}, L.~{Fletcher}, C.~J.
  {Schrijver}, D.~{Alexander}]{aschwanden_1999}
{\EM {Aschwanden} M.~J., {Fletcher} L., {Schrijver} C.~J., {Alexander} D.}
\newblock {Coronal Loop Oscillations Observed with the Transition Region and
  Coronal Explorer} \allowbreak\newblock// The Astrophysical Journal. VIII
  1999. 520. 880--894.

\bibitem[Aschwanden(2005)M.~J. Aschwanden]{2005}
{\EM Aschwanden Markus~J}.
\newblock Physics of the Solar Corona. 2005.  320.

\bibitem[Aschwanden(2006)M.~J. Aschwanden]{Aschwanden_2006}
{\EM Aschwanden Markus~J}.
\newblock Coronal magnetohydrodynamic waves and oscillations: observations and
  quests \allowbreak\newblock// Philosophical Transactions of the Royal Society
  A: Mathematical, Physical and Engineering Sciences. feb 2006. 364, 1839.
  417--432.

\bibitem[Aschwanden, Schrijver(2011)M.~J. Aschwanden, C.~J.
  Schrijver]{ashwanden2011}
{\EM Aschwanden Markus~J., Schrijver Carolus~J.}
\newblock {CORONAL} {LOOP} {OSCILLATIONS} {OBSERVED} {WITH} {ATMOSPHERIC}
  {IMAGING} {ASSEMBLY}{\textemdash}{KINK} {MODE} {WITH} {CROSS}-{SECTIONAL}
  {AND} {DENSITY} {OSCILLATIONS} \allowbreak\newblock// The Astrophysical
  Journal. jul 2011. 736, 2. 102.

\bibitem[Chen et~al.(2015)S.-X. Chen, B.~Li, L.-D. Xia, H.~Yu]{Chen_2015}
{\EM Chen Shao-Xia, Li~Bo, Xia Li-Dong, Yu~Hui}.
\newblock Periods and Damping Rates of Fast Sausage Oscillations in
  Multishelled Coronal Loops \allowbreak\newblock// Solar Physics. aug 2015.
  290, 8. 2231--2243.

\bibitem[Dadashi et~al.(2009)N.~Dadashi, H.~Safari, S.~Nasiri]{Dadashi_2009}
{\EM Dadashi N., Safari H., Nasiri S.}
\newblock The effect of the longitudinal density stratification on the standing
  kink modes for coronal loop oscillations \allowbreak\newblock// Iranian
  Journal of Physics Research. jan 2009. 9, 3. 227--235.

\bibitem[Dymova, Ruderman(2005)M.~V. Dymova, M.~S. Ruderman]{Dymova_2005}
{\EM Dymova M.~V., Ruderman M.~S.}
\newblock Non-Axisymmetric Oscillations of Thin Prominence Fibrils
  \allowbreak\newblock// Solar Physics. jun 2005. 229, 1. 79--94.

\bibitem[{Edwin}, {Roberts}(1983)P.~M. {Edwin}, B.~{Roberts}]{edwin_1983}
{\EM {Edwin} P.~M., {Roberts} B.}
\newblock {Wave propagation in a magnetic cylinder} \allowbreak\newblock//
  Solar Physics. X 1983. 88. 179--191.

\bibitem[Erdelyi, Fedun(2007)R.~Erdelyi, V.~Fedun]{Erdelyi_2007}
{\EM Erdelyi R., Fedun V.}
\newblock Are There Alfven Waves in the Solar Atmosphere?
  \allowbreak\newblock// Science. dec 2007. 318, 5856. 1572--1574.

\bibitem[Erd{\'{e}}lyi, Taroyan(2008)R.~Erd{\'{e}}lyi, Y.~Taroyan]{erdelyi2008}
{\EM Erd{\'{e}}lyi R., Taroyan Y.}
\newblock Hinode {EUV} spectroscopic observations of coronal oscillations
  \allowbreak\newblock// Astronomy {\&} Astrophysics. sep 2008. 489, 3.
  L49--L52.

\bibitem[{Esmaeili} et~al.(2015)S.~{Esmaeili}, M.~{Nasiri}, N.~{Dadashi},
  H.~{Safari}]{Esmaeili_2015}
{\EM {Esmaeili} S., {Nasiri} M., {Dadashi} N., {Safari} H.}
\newblock {Eigen-Frequencies of MHD Wave Equations in the Presence of
  Longitudinal Stratification Density} \allowbreak\newblock// AAS/AGU Triennial
  Earth-Sun Summit.  1. IV 2015.  403.17.
\newblock (AAS/AGU Triennial Earth-Sun Summit).

\bibitem[{Esmaeili} et~al.(2017)S.~{Esmaeili}, M.~{Nasiri}, N.~{Dadashi},
  H.~{Safari}]{Esmaeili_2017}
{\EM {Esmaeili} S., {Nasiri} M., {Dadashi} N., {Safari} H.}
\newblock {Behavior of Eigenfrequencies in a System of Coronal Loops
  Oscillation: Multi-stranded Loops Interaction Approach}
  \allowbreak\newblock// SOLARNET IV: The Physics of the Sun from the Interior
  to the Outer Atmosphere. I 2017.  116.

\bibitem[Esmaeili et~al.(2016)S.~Esmaeili, M.~Nasiri, N.~Dadashi,
  H.~Safari]{Esmaeili_2016}
{\EM Esmaeili Shahriar, Nasiri Mojtaba, Dadashi Neda, Safari Hossein}.
\newblock Wave function properties of a single and a system of magnetic flux
  tube(s) oscillations \allowbreak\newblock// Journal of Geophysical Research:
  Space Physics. oct 2016. 121, 10. 9340--9355.

\bibitem[Farahani et~al.(2017)S.~V. Farahani, E.~Ghanbari, G.~Ghaffari,
  H.~Safari]{Vasheghani_Farahani_2017}
{\EM Farahani S.~Vasheghani, Ghanbari E., Ghaffari G., Safari H.}
\newblock Torsional wave propagation in solar tornadoes \allowbreak\newblock//
  Astronomy {\&} Astrophysics. feb 2017. 599. A19.

\bibitem[Farahani et~al.(2014)S.~V. Farahani, C.~Hornsey, T.~van Doorsselaere,
  M.~Goossens]{Farahani_2014}
{\EM Farahani S.~Vasheghani, Hornsey C., Doorsselaere T. van, Goossens M.}
\newblock {FREQUENCY} {AND} {DAMPING} {RATE} {OF} {FAST} {SAUSAGE} {WAVES}
  \allowbreak\newblock// The Astrophysical Journal. jan 2014. 781, 2. 92.

\bibitem[Fathalian, Safari(2010)N.~Fathalian, H.~Safari]{Fathalian_2010}
{\EM Fathalian N., Safari H.}
\newblock {TRANSVERSE} {OSCILLATIONS} {OF} A {LONGITUDINALLY} {STRATIFIED}
  {CORONAL} {LOOP} {SYSTEM} \allowbreak\newblock// The Astrophysical Journal.
  nov 2010. 724, 1. 411--416.

\bibitem[Goossens et~al.(2006)M.~Goossens, J.~Andries,
  I.~Arregui]{Goossens_2006}
{\EM Goossens M, Andries J, Arregui I}.
\newblock Damping of magnetohydrodynamic waves by resonant absorption in the
  solar atmosphere \allowbreak\newblock// Philosophical Transactions of the
  Royal Society A: Mathematical, Physical and Engineering Sciences. feb 2006.
  364, 1839. 433--446.

\bibitem[Goossens et~al.(2002)M.~Goossens, J.~Andries, M.~J.
  Aschwanden]{Goossens_2002}
{\EM Goossens M., Andries J., Aschwanden M.~J.}
\newblock Coronal loop oscillations \allowbreak\newblock// Astronomy {\&}
  Astrophysics. oct 2002. 394, 3. L39--L42.

\bibitem[Goossens et~al.(1992)M.~Goossens, J.~V. Hollweg,
  T.~Sakurai]{Goossens_1992}
{\EM Goossens Marcel, Hollweg Joseph~V., Sakurai Takashi}.
\newblock Resonant behaviour of {MHD} waves on magnetic flux tubes
  \allowbreak\newblock// Solar Physics. apr 1992. 138, 2. 233--255.

\bibitem[Goossens et~al.(1995)M.~Goossens, M.~S. Ruderman, J.~V.
  Hollweg]{Goossens_1995}
{\EM Goossens Marcel, Ruderman Michail~S., Hollweg Joseph~V.}
\newblock Dissipative {MHD} solutions for resonant Alfven waves in
  1-dimensional magnetic flux tubes \allowbreak\newblock// Solar Physics. mar
  1995. 157, 1-2. 75--102.

\bibitem[Grant et~al.(2015)S.~D.~T. Grant, D.~B. Jess, M.~G. Moreels, R.~J.
  Morton, D.~J. Christian, I.~Giagkiozis, G.~Verth, V.~Fedun, P.~H. Keys,
  T.~van Doorsselaere, R.~Erd{\'{e}}lyi]{Grant_2015}
{\EM Grant S.~D.~T., Jess D.~B., Moreels M.~G., Morton R.~J., Christian D.~J.,
  Giagkiozis I., Verth G., Fedun V., Keys P.~H., Doorsselaere T. van,
  Erd{\'{e}}lyi R.}
\newblock {WAVE} {DAMPING} {OBSERVED} {IN} {UPWARDLY} {PROPAGATING}
  {SAUSAGE}-{MODE} {OSCILLATIONS} {CONTAINED} {WITHIN} A {MAGNETIC} {PORE}
  \allowbreak\newblock// The Astrophysical Journal. jun 2015. 806, 1. 132.

\bibitem[Guo et~al.(2015)Y.~Guo, R.~Erd{\'{e}}lyi, A.~K. Srivastava, Q.~Hao,
  X.~Cheng, P.~F. Chen, M.~D. Ding, B.~N. Dwivedi]{Guo_2015}
{\EM Guo Y., Erd{\'{e}}lyi R., Srivastava A.~K., Hao Q., Cheng X., Chen P.~F.,
  Ding M.~D., Dwivedi B.~N.}
\newblock {MAGNETOHYDRODYNAMIC} {SEISMOLOGY} {OF} A {CORONAL} {LOOP} {SYSTEM}
  {BY} {THE} {FIRST} {TWO} {MODES} {OF} {STANDING} {KINK} {WAVES}
  \allowbreak\newblock// The Astrophysical Journal. jan 2015. 799, 2. 151.

\bibitem[Hollweg, Yang(1988)J.~V. Hollweg, G.~Yang]{Hollweg_1988}
{\EM Hollweg Joseph~V., Yang G.}
\newblock Resonance absorption of compressible magnetohydrodynamic waves at
  thin {\textquotedblleft}surfaces{\textquotedblright} \allowbreak\newblock//
  Journal of Geophysical Research. 1988. 93, A6. 5423.

\bibitem[Jin et~al.(2018)Y.-J. Jin, H.-N. Zheng, Z.-P. Su]{Jin_2018}
{\EM Jin Yong-Jian, Zheng Hui-Nan, Su~Zhen-Peng}.
\newblock Propagation and Damping of Two-Fluid Magnetohydrodynamic Waves in
  Stratified Solar Atmosphere \allowbreak\newblock// Chinese Physics Letters.
  jul 2018. 35, 7. 075201.

\bibitem[Karami et~al.(2002)K.~Karami, S.~Nasiri, Y.~Sobouti]{Karami_2002}
{\EM Karami K., Nasiri S., Sobouti Y.}
\newblock Normal modes of magnetic flux tubes and dissipation
  \allowbreak\newblock// Astronomy {\&} Astrophysics. dec 2002. 396, 3.
  993--1002.

\bibitem[Li et~al.(2018)D.~Li, D.~Yuan, Y.~N. Su, Q.~M. Zhang, W.~Su, Z.~J.
  Ning]{Li_2018}
{\EM Li~D., Yuan D., Su~Y.~N., Zhang Q.~M., Su~W., Ning Z.~J.}
\newblock Non-damping oscillations at flaring loops \allowbreak\newblock//
  Astronomy {\&} Astrophysics. sep 2018. 617. A86.

\bibitem[Mendoza-Briceno et~al.(2004)C.~A. Mendoza-Briceno, R.~Erdelyi,
  L.~D.~G. Sigalotti]{Mendoza_Briceno_2004}
{\EM Mendoza-Briceno Cesar~A., Erdelyi Robert, Sigalotti Leonardo Di~G.}
\newblock The Effects of Stratification on Oscillating Coronal Loops
  \allowbreak\newblock// The Astrophysical Journal. apr 2004. 605, 1. 493--502.

\bibitem[Moortel, Brady(2007)I.~D. Moortel, C.~S. Brady]{De_Moortel_2007}
{\EM Moortel I.~De, Brady C.~S.}
\newblock Observation of Higher Harmonic Coronal Loop Oscillations
  \allowbreak\newblock// The Astrophysical Journal. aug 2007. 664, 2.
  1210--1213.

\bibitem[Nakariakov(1999)V.~M. Nakariakov]{Nakariakov_1999}
{\EM Nakariakov V.~M.}
\newblock {TRACE} Observation of Damped Coronal Loop Oscillations: Implications
  for Coronal Heating \allowbreak\newblock// Science. aug 1999. 285, 5429.
  862--864.

\bibitem[Ofman, Aschwanden(2002)L.~Ofman, M.~J. Aschwanden]{Ofman_2002}
{\EM Ofman L., Aschwanden M.~J.}
\newblock Damping Time Scaling of Coronal Loop Oscillations Deduced from
  [{ITAL}]Transition Region and Coronal Explorer[/{ITAL}] Observations
  \allowbreak\newblock// The Astrophysical Journal. sep 2002. 576, 2.
  L153--L156.

\bibitem[Ofman, Wang(2002)L.~Ofman, T.~Wang]{Ofman_tong_2002}
{\EM Ofman L., Wang Tongjiang}.
\newblock Hot Coronal Loop Oscillations Observed by {SUMER}: Slow Magnetosonic
  Wave Damping by Thermal Conduction \allowbreak\newblock// The Astrophysical
  Journal. nov 2002. 580, 1. L85--L88.

\bibitem[Pascoe et~al.(2018)D.~J. Pascoe, S.~A. Anfinogentov, C.~R. Goddard,
  V.~M. Nakariakov]{Pascoe_2018}
{\EM Pascoe D.~J., Anfinogentov S.~A., Goddard C.~R., Nakariakov V.~M.}
\newblock Spatiotemporal Analysis of Coronal Loops Using Seismology of Damped
  Kink Oscillations and Forward Modeling of {EUV} Intensity Profiles
  \allowbreak\newblock// The Astrophysical Journal. jun 2018. 860, 1. 31.

\bibitem[Pascoe et~al.(2017)D.~J. Pascoe, A.~J.~B. Russell, S.~A. Anfinogentov,
  P.~J.~A. Sim{\~{o}}es, C.~R. Goddard, V.~M. Nakariakov,
  L.~Fletcher]{Pascoe_2017}
{\EM Pascoe D.~J., Russell A.~J.~B., Anfinogentov S.~A., Sim{\~{o}}es P.~J.~A.,
  Goddard C.~R., Nakariakov V.~M., Fletcher L.}
\newblock Seismology of contracting and expanding coronal loops using damping
  of kink oscillations by mode coupling \allowbreak\newblock// Astronomy {\&}
  Astrophysics. oct 2017. 607. A8.

\bibitem[Roberts(2000)B.~Roberts]{Roberts_2000}
{\EM Roberts B.}
\newblock  \allowbreak\newblock// Solar Physics. 2000. 193, 1/2. 139--152.

\bibitem[{Roberts} et~al.(1984)B.~{Roberts}, P.~M. {Edwin}, A.~O.
  {Benz}]{roberts_1984}
{\EM {Roberts} B., {Edwin} P.~M., {Benz} A.~O.}
\newblock {On coronal oscillations} \allowbreak\newblock// The Astrophysical
  Journal. IV 1984. 279. 857--865.

\bibitem[{Ruderman}(2003)M.~S. {Ruderman}]{ruderman2003}
{\EM {Ruderman} M.~S.}
\newblock {The resonant damping of oscillations of coronal loops with elliptic
  cross-sections} \allowbreak\newblock// Astronomy {\&} Astrophysics. X 2003.
  409. 287--297.

\bibitem[Ruderman, Roberts(2002)M.~S. Ruderman, B.~Roberts]{Ruderman_2002}
{\EM Ruderman M.~S., Roberts B.}
\newblock The Damping of Coronal Loop Oscillations \allowbreak\newblock// The
  Astrophysical Journal. sep 2002. 577, 1. 475--486.

\bibitem[Safari et~al.(2006)H.~Safari, S.~Nasiri, K.~Karami,
  Y.~Sobouti]{Safari_2006}
{\EM Safari H., Nasiri S., Karami K., Sobouti Y.}
\newblock Resonant absorption in dissipative flux tubes \allowbreak\newblock//
  Astronomy {\&} Astrophysics. feb 2006. 448, 1. 375--378.

\bibitem[Safari et~al.(2007)H.~Safari, S.~Nasiri, Y.~Sobouti]{Safari_2007}
{\EM Safari H., Nasiri S., Sobouti Y.}
\newblock Fast kink modes of longitudinally stratified coronal loops
  \allowbreak\newblock// Astronomy {\&} Astrophysics. may 2007. 470, 3.
  1111--1116.

\bibitem[Sakurai et~al.(1991{\natexlab{a}})T.~Sakurai, M.~Goossens, J.~V.
  Hollweg]{Sakurai_1991}
{\EM Sakurai Takashi, Goossens Marcel, Hollweg Joseph~V.}
\newblock Resonant behaviour of {MHD} waves on magnetic flux tubes
  \allowbreak\newblock// Solar Physics. jun 1991{\natexlab{a}}. 133, 2.
  227--245.

\bibitem[Sakurai et~al.(1991{\natexlab{b}})T.~Sakurai, M.~Goossens, J.~V.
  Hollweg]{Sakurai_1991_b}
{\EM Sakurai Takashi, Goossens Marcel, Hollweg Joseph~V.}
\newblock Resonant behaviour of {MHD} waves on magnetic flux tubes
  \allowbreak\newblock// Solar Physics. jun 1991{\natexlab{b}}. 133, 2.
  247--262.

\bibitem[Schrijver et~al.(2002)C.~J. Schrijver, M.~J. Aschwanden, A.~M.
  Title]{Schrijver2002}
{\EM Schrijver Carolus~J., Aschwanden Markus~J., Title Alan~M.}
\newblock Transverse oscillations in coronal loops observed with TRACE -- I. An
  Overview of Events, Movies, and a Discussion of Common Properties and
  Required Conditions \allowbreak\newblock// Solar Physics. Mar 2002. 206, 1.
  69--98.

\bibitem[Shukhobodskiy et~al.(2018)A.~A. Shukhobodskiy, M.~S. Ruderman,
  R.~Erd{\'{e}}lyi]{Shukhobodskiy_2018}
{\EM Shukhobodskiy A.~A., Ruderman M.~S., Erd{\'{e}}lyi R.}
\newblock Resonant damping of kink oscillations of thin cooling and expanding
  coronal magnetic loops \allowbreak\newblock// Astronomy {\&} Astrophysics.
  nov 2018. 619. A173.

\bibitem[Soler et~al.(2011)R.~Soler, J.~Terradas, G.~Verth,
  M.~Goossens]{Soler_2011}
{\EM Soler R., Terradas J., Verth G., Goossens M.}
\newblock {RESONANTLY} {DAMPED} {PROPAGATING} {KINK} {WAVES} {IN}
  {LONGITUDINALLY} {STRATIFIED} {SOLAR} {WAVEGUIDES} \allowbreak\newblock// The
  Astrophysical Journal. jun 2011. 736, 1. 10.

\bibitem[Su et~al.(2018)W.~Su, Y.~Guo, R.~Erd{\'{e}}lyi, Z.~J. Ning, M.~D.
  Ding, X.~Cheng, B.~L. Tan]{Su_2018}
{\EM Su~W., Guo Y., Erd{\'{e}}lyi R., Ning Z.~J., Ding M.~D., Cheng X., Tan
  B.~L.}
\newblock Period Increase and Amplitude Distribution of Kink Oscillation of
  Coronal Loop \allowbreak\newblock// Scientific Reports. mar 2018. 8, 1.

\bibitem[Taran et~al.(2014)S.~Taran, H.~Safari, N.~Farhang]{Taran_2014}
{\EM Taran S., Safari H., Farhang N.}
\newblock Automated tracking of solar coronal loops and detection of their
  oscillations \allowbreak\newblock// Iranian Journal of Physics Research. jan
  2014. 14, 1. 65--72.

\bibitem[{Uchida}(1970)Y.~{Uchida}]{uchida}
{\EM {Uchida} Y.}
\newblock {Diagnosis of Coronal Magnetic Structure by Flare-Associated
  Hydromagnetic Disturbances} \allowbreak\newblock// Publications of the
  Astronomical Society of Japan,. 1970. 22. 341.

\bibitem[Verth et~al.(2007)G.~Verth, T.~van Doorsselaere, R.~Erd{\'{e}}lyi,
  M.~Goossens]{Verth_2007}
{\EM Verth G., Doorsselaere T. van, Erd{\'{e}}lyi R., Goossens M.}
\newblock Spatial magneto-seismology: effect of density stratification on the
  first harmonic amplitude profile of transversal coronal loop oscillations
  \allowbreak\newblock// Astronomy {\&} Astrophysics. sep 2007. 475, 1.
  341--348.

\bibitem[Verth et~al.(2010)G.~Verth, J.~Terradas, M.~Goossens]{Verth_2010}
{\EM Verth G., Terradas J., Goossens M.}
\newblock {OBSERVATIONAL} {EVIDENCE} {OF} {RESONANTLY} {DAMPED} {PROPAGATING}
  {KINK} {WAVES} {IN} {THE} {SOLAR} {CORONA} \allowbreak\newblock// The
  Astrophysical Journal. jul 2010. 718, 2. L102--L105.

\bibitem[Verwichte et~al.(2009)E.~Verwichte, M.~J. Aschwanden, T.~van
  Doorsselaere, C.~Foullon, V.~M. Nakariakov]{Verwichte_2009}
{\EM Verwichte E., Aschwanden M.~J., Doorsselaere T. van, Foullon C.,
  Nakariakov V.~M.}
\newblock {SEISMOLOGY} {OF} A {LARGE} {SOLAR} {CORONAL} {LOOP} {FROM}
  {EUVI}/{STEREOOBSERVATIONS} {OF} {ITS} {TRANSVERSE} {OSCILLATION}
  \allowbreak\newblock// The Astrophysical Journal. may 2009. 698, 1. 397--404.

\bibitem[Wang, Solanki(2004)T.~J. Wang, S.~K. Solanki]{Wang_2004}
{\EM Wang T.~J., Solanki S.~K.}
\newblock Vertical oscillations of a coronal loop observed by {TRACE}
  \allowbreak\newblock// Astronomy {\&} Astrophysics. jun 2004. 421, 2.
  L33--L36.

\bibitem[Wang et~al.(2003)T.~J. Wang, S.~K. Solanki, D.~E. Innes, W.~Curdt,
  E.~Marsch]{wang2003}
{\EM Wang T.~J., Solanki S.~K., Innes D.~E., Curdt W., Marsch E.}
\newblock Slow-mode standing waves observed by {SUMER} in hot coronal loops
  \allowbreak\newblock// Astronomy {\&} Astrophysics. apr 2003. 402, 2.
  L17--L20.

\bibitem[Zhang et~al.(2016)Q.-H. Zhang, Y.-M. Wang, R.~Liu, C.-L. Shen,
  M.~Zhang, T.-Y. Gou, J.-J. Liu, K.~Liu, Z.-J. Zhou, S.~Wang]{Zhang_2016}
{\EM Zhang Quan-Hao, Wang Yu-Ming, Liu Rui, Shen Cheng-Long, Zhang Min, Gou
  Ting-Yu, Liu Jia-Jia, Liu Kai, Zhou Zhen-Jun, Wang Shui}.
\newblock Damped large amplitude oscillations in a solar prominence and a
  bundle of coronal loops \allowbreak\newblock// Research in Astronomy and
  Astrophysics. nov 2016. 16, 11. 167.

\bibitem[{van Doorsselaere} et~al.(2004{\natexlab{a}})T.~{van Doorsselaere},
  J.~Andries, S.~Poedts, M.~Goossens]{van_Doorsselaere_2004_a}
{\EM {van Doorsselaere} T., Andries J., Poedts S., Goossens M.}
\newblock Damping of Coronal Loop Oscillations: Calculation of Resonantly
  Damped Kink Oscillations of One-dimensional Nonuniform Loops
  \allowbreak\newblock// The Astrophysical Journal. may 2004{\natexlab{a}}.
  606, 2. 1223--1232.

\bibitem[{van Doorsselaere} et~al.(2004{\natexlab{b}})T.~{van Doorsselaere},
  A.~{Debosscher}, J.~{Andries}, S.~{Poedts}]{van_Doorsseleare_2004_b}
{\EM {van Doorsselaere} T., {Debosscher} A., {Andries} J., {Poedts} S.}
\newblock {The effect of curvature on quasi-modes in coronal loops}
  \allowbreak\newblock// Astronomy {\&} Astrophysics. IX 2004{\natexlab{b}}.
  424. 1065--1074.

\end{thebibliography}
\end{document}